\documentstyle[preprint,prc,aps,psfig]{revtex}
\begin{document}
\title{$NN$ interaction in a Goldstone boson exchange model}
\author{D. Bartz\thanks{e-mail : d.bartz@ulg.ac.be} and Fl.
Stancu\thanks{e-mail
: fstancu@ulg.ac.be}}
\address{Universit\'{e} de Li\`ege, Institut de Physique B.5, Sart Tilman,
B-4000 Li\`ege 1, Belgium}
\date{\today}
\maketitle
\everymath={\displaystyle}

\vspace{1cm}

\begin{abstract} Adiabatic nucleon-nucleon potentials are calculated in a
six-quark nonrelativistic chiral constituent quark model where the Hamiltonian
contains a linear confinement and a pseudoscalar meson  
(Goldstone boson) exchange interaction
between quarks.
Calculations are performed both in a cluster model and a molecular
orbital basis, through coupled channels. 
In both cases the potentials present an important
hard core at short distances, explained through the dominance of the
$[51]_{FS}$ configuration, but do not exhibit an attractive pocket.
We add a scalar 
meson exchange interaction and show how it can account for some 
middle-range attraction.
\end{abstract}

\section{Introduction}
There have been many attempts to study the nucleon-nucleon interaction starting
from a system of six interacting quarks described by a constituent quark model.
These models explain the short range repulsion as due to the colour magnetic
part of the one gluon exchange (OGE) interaction between quarks and due to
quark interchanges between two $3q$ clusters \cite{OK84,SH89}. To the OGE
interaction it was necessary to add a scalar and a pseudoscalar meson exchange
interaction
between quarks of different $3q$ clusters in order to explain the intermediate-
and long-range attraction between two nucleons \cite{KU91,ZH94,FU96}.\par
In a previous work \cite{BA98} we have calculated the nucleon-nucleon ($NN$)
interaction potential at zero-separation distance between two three-quark
clusters in the frame of a constituent quark model \cite{GL96a,GL96b,GL97a}
where the quarks interact via pseudoscalar meson i.e.
Goldstone boson exchange (GBE) instead of OGE. An
important motivation in using the GBE model is that it describes well the
baryon spectra. In particular, it correctly reproduces the order of positive
and negative parity states both for nonstrange \cite{GL96b} and strange
\cite{GL97a} baryons where the OGE model has failed.\par
The underlying symmetry of
the GBE model is related to the flavour-spin $SU_F(3) \times SU_S(2)$ group.
Combining it with the $S_3$ symmetry, a thorough analysis performed for the
$L=1$ baryons \cite{CO98} has shown that the chiral quark picture leads 
to more satisfactory fits to the observed baryon spectrum than the OGE
models.\par
The one-pion exchange potential between quarks appears naturally as an
iteration of the instanton induced interaction in the t-channel \cite{GV99}.
The meson exchange picture is also supported by explicit QCD
latice calculations \cite{LIU98}.\par 
Another motivation in using the GBE model is that the exchange interaction
contains the basic ingredients required by the $NN$ problem. Its long-range
part, required to provide the long-range $NN$ interaction,
is a Yukawa-type potential depending on the mass of the exchange meson,
Its short-range part, of
opposite sign to the long-range one, is mainly responsible for the good
description of the baryon spectra \cite{GL96a,GL96b,GL97a} and also induces a
short-range repulsion in the $NN$ system, both in the $^3S_1$ and $^1S_0$
channels \cite{ST97}. The present study is an extention of \cite{BA98} and 
we calculate here the interaction potential
between two $3q$ clusters as a function of $Z$, the separation distance between
the centres of the clusters. This separation distance is a good
approximation of the Jacobi relative coordinate between the two clusters. Under
this assumption, here we calculate the interaction potential in the adiabatic
(Born-Oppenheimer) approximation, as explained below.\par
A common issue in solving the $NN$ problem is the construction of adequate
six-quark basis states. The usual choice is a cluster model basis
\cite{OK84,SH89,HA81}. In calculating the potential at zero-separation
distance,
in Ref. \cite{BA98} we used molecular-type orbitals \cite{ST87} and
compared the
results with those based on cluster model single-particle states. The molecular
orbitals have the proper axially and reflectionally symmetries and can be
constructed from appropriate combinations of two-centre Gaussians. At
zero-separation between the $3q$ clusters the six-quark states obtained
from such orbitals contain certain $p^ns^{6-n}$ 
configurations which are missing in the
cluster model basis. By using molecular orbitals, in Ref. \cite{BA98} we found
that the height of the repulsion reduces by about 22 \% and 25 \% in the
$^3S_1$ and $^1S_0$ channels respectively with respect to cluster model
results.
It is therefore useful to analyse the role of molecular orbitals at distances
$Z \ne 0$. By construction, at $Z \rightarrow \infty$ the molecular orbital
states are simple parity conserving linear combinations of cluster model
states. Their role is expected to be important at short 
range at least. They also have the advantage of forming an orthogonal and
complete basis while the cluster model (two-centre) states are not orthogonal
and are overcomplete. For this reason we found that in practice they are more
convenient to be used than the cluster model basis, where one must carefully
\cite{HA81} consider the limit $Z \rightarrow 0$.  Here too, for the purpose of
comparison we perform calculations both in the cluster model and the molecular
orbital basis.\par
In Sec. 2 we recall the procedure of constructing molecular orbital
single-particle states starting from the two-centre Gaussians used in the
cluster model calculations. In Sec. 3 the GBE Hamiltonian is presented. Sec. 4
is devoted to the results obtained for the $NN$ potential.
In Sec. 5 we introduce a middle range attraction through a scalar
meson exchange interaction between quarks parametrized consistently
with the pseudoscalar meson exchange. The last section is
devoted to a summary and conclusions.
\section{The single-particle orbitals}
In the cluster model 
one can define states which in the limit of large
intercluster separation $Z$ are right $R$ and left $L$ states
\begin{equation}
R = \psi \left(\vec{r} - \frac{\vec{Z}}{2}\right) \hspace{8mm} \mbox{and}
\hspace{8mm} L =
\psi \left(\vec{r} + \frac{\vec{Z}}{2}\right).
\end{equation}
In the simplest cluster model basis these are ground state harmonic oscillator
wave functions centered at $Z/2$ and $- Z/2$ respectively. They contain a
parameter $\beta$ which is fixed variationally to minimize the nucleon mass
described as a $3q$ cluster within a given Hamiltonian. The states (1) are
normalized but are not orthogonal at finite $Z$. They have good parity about
their centers but not about their common center $\vec{r} = 0$.\par
From $R$ and $L$ one constructs six-quark states of given orbital symmetry
$[f]_O$. The totally antisymmetric six-quark states also contain a flavour-spin
part of symmetry $[f]_{FS}$ and a colour part of symmetry $[222]_C$. In the
cluster model the most important basis states \cite{ST97} for the Hamiltonian
described in the following section are
\begin{equation}
\left|{\left.{{R}^{3}{L}^{3}\ {\left[{6}\right]}_{O}\
{\left[{33}\right]}_{FS}}\right\rangle}\right.
\end{equation}
\begin{equation}
\left|{\left.{{R}^{3}{L}^{3}\ {\left[{42}\right]}_{O}\
{\left[{33}\right]}_{FS}}\right\rangle}\right.
\end{equation}
\begin{equation}
\left|{\left.{{R}^{3}{L}^{3}\ {\left[{42}\right]}_{O}\
{\left[{51}\right]}_{FS}}\right\rangle}\right.
\end{equation}
\begin{equation}
\left|{\left.{{R}^{3}{L}^{3}\ {\left[{42}\right]}_{O}\
{\left[{411}\right]}_{FS}}\right\rangle}\right.
\end{equation}
Harvey \cite{HA81} has shown that with a proper normalization the symmetry
$[6]_O$ contains only $s^6$ and $[42]_O$ only $s^4p^2$ configurations in the
limit $Z \rightarrow 0$.\par
According to Ref. \cite{ST87} let us consider now molecular orbital
single-particle states. 
Most generally these are eigenstates of a Hamiltonian $H_0$ having
axial and reflectional symmetries characteristic to the $NN$ problem. These
eigenstates have therefore good parity and good angular momentum projection. As
in the cluster model basis where one uses the two lowest states $R$ and $L$, in
the molecular orbital basis we also consider the two lowest states, $\sigma$,
of positive parity and $\pi$, of negative parity. From these we can construct
pseudo-right $r$ and pseudo-left $l$ states as
\begin{equation}
\left[ \begin{array}{c} r\\ l \end{array} \right] =
2^{-1/2}\, ( \sigma \pm \pi ) \hspace{5mm}
\rm{for\ all\ }Z,
\end{equation}
where
\begin{eqnarray}
& <r|r> = <l|l> = 1 ,\,\, <r|l> = 0 .&
\end{eqnarray}
In principle one can obtain molecular orbital single particle states from mean
field calculations (see for example \cite{KO94}). Here we approximate them by good
parity, orthonormal states constructed from the cluster model states (1) as
\begin{equation}
\left[ \begin{array}{c} \sigma\\ \pi\end{array} \right] =
[2( 1 \pm <R|L>)]^{-1/2} ( R \pm L ) ,
\end{equation}
Such molecular orbitals are a very good approximation to the exact eigenstates
of a ``two-centre" oscillator frequently used in nuclear physics or
occasionally \cite{RO87} in the calculation of the $NN$ potential. They provide
a convenient basis for the first step calculations based on the adiabatic
approximation as described below.\par
Introduced in (6) they give
\begin{equation}
\left[ \begin{array}{c} r\\ l \end{array} \right] =
\frac{1}{2} \left[ \frac{R+L}{(1+<R|L>)^{1/2}} \pm
\frac{R-L}{(1-<R|L>)^{1/2}} \right].
\end{equation}
At $Z \rightarrow 0$ one has $\sigma \rightarrow s$ and $\pi \rightarrow p$
(with $m = 0, \pm 1$) where $s$ and $p$ are harmonic oscillator states. Thus in
the limit $Z \rightarrow 0$ one has
\begin{equation}
\left[ \begin{array}{c} r\\ l \end{array} \right] =
2^{1/2} (s \pm p) ,
\end{equation}
and at $Z \rightarrow \infty$ one recovers the cluster model basis because $r
\rightarrow R$ and $\ell \rightarrow L$.
\par Equation (9) with $R$ and $L$ defined by (1) ensures that the same $Z$
is used both in the molecular and the cluster model basis.\par
From $(r,l)$ as well as from $(\sigma,\pi)$ orbitals one can construct
six-quark states of required permutation symmetry. For the $S_6$
symmetries relevant for the $NN$ problem the transformations
between six-quark states expressed in terms of $(r,l)$ and $(\sigma,\pi)$
states are given in Table I of Ref. \cite{ST87}. This table shows that in
the limit $Z \rightarrow 0$ six-quark states obtained from molecular
orbitals contain configurations of type $s^np^{6-n}$ with $n = 0,1,...,
6$. For example the $[6]_O$ state contains $s^6$,
$s^6p^4$, $s^2p^4$ and $p^6$ configurations and the $[42]_O$ state
associated to
the $S$-channel contains $s^4p^2$ and $s^2p^4$ configurations. This is in
contrast to the cluster model basis where $[6]_O$ contains only $s^6$ and
$[42]_O$ only $s^4p^2$ configurations, as mentioned above. This suggests that 
the six-quark basis states constructed from molecular orbitals form a richer
basis without introducing more single particle states.

Using Table I of Ref. \cite{ST87} we find that the six-quark basis states
needed for the $^3S_1$ or $^1S_0$ channels are:
\begin{eqnarray}
\left.{\left|{33{\left[{6}\right]}_{O}{\left[{33}\right]}_{FS}}\right.}
\right\rangle\
& = & \frac{1}{4}\ \left.{\left| \left[{\sqrt {5}\ \left({{\sigma}^{6}\ -\
{\pi}^{6}}\right)\ -\
\sqrt {3}\ \left({{\sigma}^{4}{\pi}^{2}\ -\ {\sigma}^{2}{\pi}^{4}}\right)}\right]\
{{\left[{6}\right]}_{O}{\left[{33}\right]}_{FS}}\right.}\right\rangle, \\
\left.{\left|{33{\left[{42}\right]}_{O}{\left[{33}\right]}_{FS}}\right.}
\right\rangle\
& = & \sqrt {\frac{1}{2}}\ \left.{\left|{ \left[{{\sigma}^{4}{\pi}^{2}\ -\
{\sigma}^{2}{\pi}^{4}}\right]
{\left[{42}\right]}_{O}{\left[{33}\right]}_{FS}}\right.}\right\rangle, \\
\left.{\left|{33{\left[{42}\right]}_{O}{\left[{51}\right]}_{FS}}\right.}
\right\rangle\
& = & \sqrt {\frac{1}{2}}\ \left.{\left|{ \left[{{\sigma}^{4}{\pi}^{2}\ -\
{\sigma}^{2}{\pi}^{4}}\right]
{\left[{42}\right]}_{O}{\left[{51}\right]}_{FS}}\right.}\right\rangle, \\
\left.{\left|{33{\left[{42}\right]}_{O}{\left[{411}\right]}_{FS}}\right.}
\right\rangle\
& = & \sqrt {\frac{1}{2}}\ \left.{\left|{ \left[{{\sigma}^{4}{\pi}^{2}\ -\
{\sigma}^{2}{\pi}^{4}}\right]
{\left[{42}\right]}_{O}{\left[{411}\right]}_{FS}}\right.}\right\rangle, \\
\left.{\left|{{42}^{+}{\left[{6}\right]}_{O}{\left[{33}\right]}_{FS}}\right.}
\right\rangle\
& = & {\frac{1}{4}} \sqrt {{\frac{1}{2}}}\ \left.{\left|{ \left[{{\sqrt
{15}}^{}\left({{\sigma}^{6}\ +\ {\pi}^{6}}\right)\
 -\ \left({{\sigma}^{4}{\pi}^{2}\ +\
{\sigma}^{2}{\pi}^{4}}\right)}\right]{\left[{6}\right]}_{O}{\left[{33}\right]}_{FS}}
\right.}\right\rangle, \\
\left.{\left|{{42}^{+}{\left[{42}\right]}_{O}{\left[{33}\right]}_{FS}}\right.}
\right\rangle\
& = & \sqrt {\frac{1}{2}}\ \left.{\left|{ \left[{\left.{{\sigma}^{4}
{\pi}^{2}}\right.\ +\
{\sigma}^{2}{\pi}^{4}}\right]{\left[{42}\right]}_{O}{\left[{33}\right]}_{FS}}\right.}
\right\rangle, \\
\left.{\left|{{42}^{+}{\left[{42}\right]}_{O}{\left[{51}\right]}_{FS}}\right.}
\right\rangle\
& = & \sqrt {\frac{1}{2}}\ \left.{\left|{ \left[{\left.{{\sigma}^{4}
{\pi}^{2}}\right.\ +\
{\sigma}^{2}{\pi}^{4}}\right]{\left[{42}\right]}_{O}{\left[{51}\right]}_{FS}}\right.}
\right\rangle, \\
\left.{\left|{{42}^{+}{\left[{42}\right]}_{O}{\left[{411}\right]}_{FS}}\right.}
\right\rangle\
& = & \sqrt {\frac{1}{2}}\ \left.{\left|{ \left[{\left.{{\sigma}^{4}
{\pi}^{2}}\right.\ +\
{\sigma}^{2}{\pi}^{4}}\right]{\left[{42}\right]}_{O}{\left[{411}\right]}_{FS}}\right.}
\right\rangle, \\
\left.{\left|{{51}^{+}{\left[{6}\right]}_{O}{\left[{33}\right]}_{FS}}\right.}
\right\rangle\
& = & \frac{1}{4}\ \left.{\left|{ \left[{\sqrt {3}\ \left.{\left({{\sigma}^{6}\ - \
{\pi}^{6}}\right)}\right.\ +\ \sqrt {5}\ \left({{\sigma}^{4}{\pi}^{2}\ -\
{\sigma}^{2}{\pi}^{4}}\right)}\right]{\left[{6}\right]}_{O}{\left[{33}\right]}_{FS}}
\right.}\right\rangle,
\end{eqnarray}
\noindent
where the notation $33$ and $mn^+$ in the left-hand side of each equality above 
means $r^3\ell^3$ and $r^m\ell^n+r^n\ell^m$ as in Ref. \cite{ST87}. 
Each wave function contains an orbital part ($O$) and a
flavour-spin part ($FS$) which combined with the colour singlet $[222]_C$ state
gives rise to a totally antisymmetric state. We restricted the flavour-spin
states to $[33]_{FS}$, $[51]_{FS}$ and $[411]_{FS}$ as for the cluster model
basis (2-5).\par

As explained above, besides being poorer in $s^np^{6-n}$ configurations, the
number of basis states is smaller in the cluster model although we deal
with the
same $[f]_O$ and $[f]_{FS}$ symmetries and the same harmonic oscillator states
$s$ and $p$ in both cases. This is due to the existence of three-quark clusters
only
in the cluster model states, while the molecular basis also allows
configurations with five quarks to the left and one to the right, or vice
versa, or four quarks to the left and two to the right or vice versa. At large
separations these states act as ``hidden colour" states but at 
short- and medium-range separation distances they are expected to
bring a significant contribution, as we shall see below.
The ''hidden colour" are states where a $3q$ cluster in an $s^3$ configuration
is a colour octet, in contrast to the nucleon which is a colour
singlet. Their role is important at short separations but it
vanishes at large ones
(see e.g. \cite{HA81}).\par

\section{Hamiltonian}
The GBE Hamiltonian considered in this study has the form \cite{GL96b,GL97a} :
\begin{equation}
H= \sum_i m_i + \sum_i \frac{\vec{p}_{i}^{\,2}}{2m_i} - \frac {(\sum_i
\vec{p}_{i})^2}{2\sum_i m_i} + \sum_{i<j} V_{\text{conf}}(r_{ij}) + \sum_{i<j}
V_\chi(r_{ij}) \, ,
\label{ham}
\end{equation}
with the linear confining interaction :
\begin{equation}
 V_{\text{conf}}(r_{ij}) = -\frac{3}{8}\lambda_{i}^{c}\cdot\lambda_{j}^{c} 
\, ( V_0 +C\, r_{ij} \,) ,
\label{conf}
\end{equation}
and the spin--spin component of the GBE interaction in its $SU_F(3)$ form :
\begin{eqnarray}
V_\chi(r_{ij})
&=&
\left\{\sum_{F=1}^3 V_{\pi}(r_{ij}) \lambda_i^F \lambda_j^F \right.
\nonumber \\
&+& \left. \sum_{F=4}^7 V_{K}(r_{ij}) \lambda_i^F \lambda_j^F
+V_{\eta}(r_{ij}) \lambda_i^8 \lambda_j^8
+V_{\eta^{\prime}}(r_{ij}) \lambda_i^0 \lambda_j^0\right\}
\vec\sigma_i\cdot\vec\sigma_j,
\label{VCHI}
\end{eqnarray}
\noindent
with $\lambda^0 = \sqrt{2/3}~{\bf 1}$, where $\bf 1$ is the $3\times3$ unit
matrix. The interaction (22) contains $\gamma = \pi, K, \eta$ and $\eta '$
meson-exchange terms and the form of $V_{\gamma} \left(r_{ij}\right)$ is given
as the sum of two distinct contributions : a Yukawa-type potential containing
the mass of the exchanged meson and a short-range contribution of opposite
sign, the role of which is crucial in baryon spectroscopy.\par

In the parametrization of Ref. \cite{GL96b} the exchange potential
due to a meson $\gamma$ has the form
\begin{equation}V_\gamma (r)=
\frac{g_\gamma^2}{4\pi}\frac{1}{12m_i m_j}
\{\theta(r-r_0)\mu_\gamma^2 \,\frac{e^{-\mu_\gamma r}}{ r}- \frac {4}{\sqrt {\pi}}
\alpha^3 \exp(-\alpha^2(r-r_0)^2)\}.
\end{equation}
The shifted Gaussian of Eq. (23) results from a pure phenomenological fit
(see below) of the baryon spectrum with
\begin{equation}
r_0 = 0.43 \, { fm}, ~\alpha = 2.91 \, { fm}^{-1},~~
\end{equation}
\par

For a system of $u$ and $d$ quarks only, as it is the case here,
the $K$-exchange does not contribute. The apriori determined
parameters of the GBE model are the masses
\begin{equation}
m_{u,d} = 340 \, { MeV}, \,
\mu_{\pi} = 139 \, { MeV},~ \mu_{\eta} = 547 \, { MeV},~
\mu_{\eta'} = 958 \, { MeV}.
\end{equation}
The other parameters are given in Table I.

It is useful to  comment on Eq. (23). The coupling of pseudoscalar mesons to 
quarks (or nucleons) gives rise to a two-body interaction potential which 
contains a Yukawa-type term and a contact term of opposite sign (see e.g. 
\cite{BJ76}).  
The second term of (23) stems from the contact term, regularized with 
parameters fixed phenomenologically.  Certainly more fundamental studies 
are required to understand this second term and attempts are being made 
in this direction.  The instanton liquid model of the vacuum (for a review 
see \cite{SS98}) implies point-like quark-quark 
interactions.  To obtain a realistic description of the hyperfine 
interaction this interaction has to be iterated in the t-channel \cite{GV99}.  
The t-channel iteration admits a meson exchange interpretation \cite{RB99}.

In principle it would be better to use a parametrization of the GBE
interaction as given in \cite{GL97b} based on a semirelativistic
Hamiltonian. However, in applying the quark cluster approach to
two-baryon systems we are restricted to use a nonrelativistic
kinematics and an $s^3$ wave function for the ground state baryon.

The matrix elements of the Hamiltonian (20) are calculated in the bases
(2)-(5) and (11)-(19)
by using the fractional parentage technique described in Refs. \cite{HA81,ST96}
and also applied in Ref. \cite{ST97}. A programme based on Mathematica
\cite{WO97} has been
created for this purpose. In this way every six-body matrix element reduces
to a
linear combination of two-body matrix elements of either symmetric or
antisymmetric states for which Eqs. (3.3) of Ref. \cite{GL96a} can be used to
integrate in the flavour-spin space.
\section{Results}
We diagonalize the Hamiltonian (20)-(25) in the six-quark cluster model basis
(2)-(5) and in the six-quark molecular orbital basis (11)-(19) for values of the
separation distance $Z$ up to 2.5 fm. Using in each case the lowest eigenvalue,
denoted by $\langle H\rangle_Z$ we define the $NN$ interaction potential in the
adiabatic (Born-Oppenheimer) approximation as
\begin{equation}
V_{NN}\left(Z\right) = \langle H\rangle_Z - 2m_N - K_{rel}
\end{equation}
Here $m_N$ is the nucleon mass obtained as a variational $s^3$ solution for a
$3q$ system described by the Hamiltonian (20). The wavefunction has the form
$\phi  \ \propto  \ \exp\ \ \left[{-\ \left({{\rho }^{2}\ +\ {\lambda
}^{2}}\right)/2{\beta }^{2}}\right]$ where $\rho =
\left(\vec{r}_1-\vec{r}_2\right)/\sqrt{2}$ and $\vec{\lambda} =
\left(\vec{r}_1+\vec{r}_2-2\vec{r}_3\right)/\sqrt{6}$. The variational solution
for $m_N = \langle H\rangle_{3q}$ and the corresponding $\beta$ is given in
Table II. The same value of $\beta$ is also used for the 6q system. This
is equivalent with imposing the ``stability condition" which is of crucial
importance in resonating group method (RGM) calculations \cite{OK84,SH89}. The
quantity $K_{rel}$ represents the relative kinetic energy of two $3q$ clusters
separated at infinity
\begin{equation}
{K}_{rel}\ =\ {\frac{{3\hbar }^{2}}{4{m\beta }^{2}}}\ 
\end{equation}
where $m$ above and in the following designates the mass of the
$u$ or $d$ quark.
For the value of $\beta$ of Table II this gives $K_{rel} = 0.448$ GeV.
\subsection{Cluster model}

In Fig. 1 we present the expectation value of the kinetic
energy $\langle KE\rangle$ as a function of $Z$. One can see that for the state 
$\left|{\left.{{R}^{3}{L}^{3}{\left[{42}\right]}_{O}}\right\rangle}\right.$
it decreases with $Z$ but for the state
$\left|{\left.{{R}^{3}{L}^{3}{\left[{6}\right]}_{O}}\right\rangle}\right.$
it first reaches a minimum at
around $Z \cong$ 0.85 fm and then tends to an asymptotic value equal to its
value at the origin due to its $s^6$ structure. This value is
\begin{equation}
\langle KE\rangle_{Z=0} = \langle KE\rangle_{Z=\infty} = \frac{15}{4} \hbar
\omega
\end{equation}
where $\hbar \omega = \hbar^2/m\beta^2$. Actually this is also the asymptotic
value for all states.\par
The diagonal matrix elements of the confinement potential are presented in Fig.
2. Beyond $Z >$ 1.5 fm one can notice a linear increase except for the
$\left|{\left.{{R}^{3}{L}^{3}{\left[{42}\right]}_{O}
{\left[{51}\right]}_{FS}}\right\rangle}\right.$ state where it reaches a
plateau
of 0.3905 GeV.\par
As an example the diagonal matrix elements of the chiral interaction $V_{\chi}$
are exhibited in Fig. 3 for $S=1$, $I=0$. At $Z=0$ one recovers the values
obtained in Ref. \cite{BA98}.
At $Z \rightarrow \infty$ the symmetries corresponding to
baryon-baryon channels, namely $[51]_{FS}$ and $[33]_{FS}$, must appear
with proper coefficients, as given by Eq. (29). 
The contribution due to these symmetries
must be identical to the contribution
of $V_{\chi}$ to two nucleon masses also calculated with the Hamiltonian
(20). This is indeed the case. In the total Hamiltonian the contribution of the
$[411]_{FS}$ $V_{\chi}$ state tends to infinity when $Z \rightarrow \infty$. Then
this state decouples from the rest which is natural because
it does not correspond
to an asymptotic baryon-baryon channel. It plays a role at small $Z$
but at large $Z$ its amplitude in the $NN$ wavefunction vanishes, similarly to
the ``hidden colour" states. Actually, in diagonalizing the total Hamiltonian
in the basis (2)-(5) we obtain an $NN$ wavefunction which in the limit $Z
\rightarrow \infty$ becomes \cite{HA81}
\begin{equation}
{\psi }_{NN}\ =\ {\frac{1}{3}}\
\left|{\left.{{\left[{6}\right]}_{O}{\left[{33}\right]}_{FS}}\right\rangle}
\right.\
+\ {\frac{2}{3}}\
\left|{\left.{{\left[{42}\right]}_{O}{\left[{33}\right]}_{FS}}\right\rangle}
\right.\
-\ {\frac{2}{3}}\
\left|{\left.{{\left[{42}\right]}_{O}{\left[{51}\right]}_{FS}}\right\rangle}
\right.
\end{equation}
The adiabatic potential drawn in Figs. 6 and
7 is defined according to Eq. (26) where
$\langle H\rangle_Z$ is the lowest eigenvalue resulting from the
diagonalization. Fig. 6 corresponds to $S$ = 1, $I$ = 0 and Fig. 7 to
$S$ = 0, $I$=1. Note that from these curves one should subtract
$K_{rel}$ of Eq. (27) in order to obtain the asymptotic value 
zero for the potential. 
One can see that the potential is repulsive at any $Z$ 
in both sectors.\par
Our cluster model results can be compared to previous literature based
on OGE models. A typical example for the $^3S_1$ and $^1S_0$ adiabatic
potentials can be found in Ref.\cite{EH84}. The results are similar to
ours. There is a repulsive core but no attractive pocket.
However, in our case, in either bases, the core is about twice higher at 
$Z$ = 0 and about 0.5 fm wider than in \cite{EH84}.

\subsection{Molecular orbital basis}
In the molecular basis
the diagonal matrix elements of the kinetic energy are similar to each other as
decreasing functions of $Z$. As an illustration in Fig. 4 we show $\langle
K.E.\rangle$ corresponding to
$\left|{\left.{33{\left[{6}\right]}_{O}{\left[{33}\right]}_{FS}}\right\rangle}
\right.$ and to the most dominant state 
at $Z=0$, namely $\left|{\left.{42^+
{\left[{42}\right]}_{O}{\left[{51}\right]}_{FS}}\right\rangle}\right.$ 
(see \cite{BA98}). 
The kinetic energy of the latter is larger than that of the former
because of the presence of the configuration $s^2 p^4$ with 50 \%
probability while in the first state this probability is smaller
as well as that of the $p^6$ configuration, see eqs. (11) and (17).
The large kinetic energy of the state (17) is compensated  
by large negative values of $\langle V_{\chi}\rangle$ 
so that this state becomes dominant at small $Z$ in agreement
with Ref. \cite{BA98}.\par 
The expectation values of the confinement potential increase with $Z$ becoming
linear beyond $Z >$ 1.5 fm except for the state
$\left|{\left.{33{\left[{42}\right]}_{O}{\left[{51}\right]}_{FS}}\right\rangle}
\right.$ which gives a result very much similar to the cluster model
state
$\left|{\left.{R^3L^3{\left[{42}\right]}_{O}{\left[{51}\right]}_{FS}}\right
\rangle}
\right.$ drawn in Fig. 2. 
Such a behaviour can be understood through the details given
in the Appendix.
Due to the similarity to the cluster model 
results we do not show here $\langle V_{conf} \rangle$ 
explicitly for the molecular orbital basis.\par
The expectation value of the chiral interaction either decreases or increases
with $Z$ depending on the state. In Fig. 5 we illustrate the case of
the
$\left|{\left.{42^+
{\left[{42}\right]}_{O}{\left[{51}\right]}_{FS}}\right\rangle}\right.$
state. both for $S$ = 1, $I$ = 0 and $S$ = 0, $I$ = 1 sectors.
This state is the dominant component of ${\psi }_{NN}$ at $Z$ = 0
with a probability of 87 \% for $SI$ = (10) and 93 \% for $SI$ = (01)
\cite{BA98}. With increasing $Z$ these probabilities decrease and
tend to zero at $Z \rightarrow \infty$. In fact in the molecular
orbital basis the asymptotic form of  ${\psi }_{NN}$ is also given
by Eq. (29) inasmuch as $r \rightarrow R$ and $l \rightarrow L$
as indicated below Eq. (10).\par

Adding together these contributions we diagonalize the Hamiltonian and use its
lowest eigenvalue to obtain the $NN$ potential according to the definition
(26). The $S=1$, $I=0$ and $S=0$, $I=1$ cases are illustrated in Figs. 6 and 7 
respectively, for a comparison with the cluster model basis. As shown in Ref.
\cite{BA98} at $Z=0$ the repulsion reduces by about 22 \% and 25 \% in the
$^3S_1$ and $^1S_0$ channels respectively when passing from the cluster model
basis to the molecular orbital basis. From Figs. 6 and 7 one can see that the
molecular orbital basis has an important effect up to about $Z \approx$ 1.5 fm
giving a lower potential 
at small values of $Z$.
For $Z \approx 1$ fm
it gives a potential larger by few tens of MeV 
than the cluster model potential.
However there is no attraction at all
in either case.\par

Actually, by construction, the molecular orbital basis 
is richer at $Z$ = 0 \cite{ST87} than the cluster model basis. For this 
reason, 
at small $Z$ it leads to a lower potential than the cluster model basis.
Within a truncated space this property may not hold beyond some value
of $Z$. However by an increase of the Hilbert space one can possibly
bring the molecular potential lower again. In fact
we chose the most important
configurations from symmetry arguments \cite{ST97} based on
Casimir operator eigenvalues. These arguments hold if the
interaction is the same for all quarks in the coordinate space.
This is certainly a better approximation for $Z$ = 0 than for 
larger values of $Z$. So it means that other configurations,
which have been neglected, may play a role at $Z > 0.4$ fm.
Then, if added, they could possibly lower the molecular basis result.
\par 

As defined in Sec. 2 the quantity $Z$ is the separation distance
between two $3q$ clusters.  It represents the Jacobi relative
coordinate between the two nucleons only for large $Z$.  There we view
it as a generator coordinate and the potential we obtain represents 
the diagonal kernel appearing in the resonating group or the generator 
coordinate method.  The comparison given above should then be considered
in the context of the generator coordinate method wich will be developped
in further studies and will lead to nonlocal potentials.  However
the adiabatic potentials, here obtained in the two bases can be compared
with each other in an independent and different way.  On can introduce
the quadrupole moment of the six-quark system
\begin{equation}
q_{20}=\sum_{i=1}^6 r_i^2\ Y_{20} (\hat{r}_i)
\end{equation}
and treat the square root of its expectation value
\begin{equation}
\langle Q\rangle=\langle \psi_{NN}|q_{20}|\psi_{NN}\rangle
\end{equation}
as a collective coordinate describing the separation between the two
nucleons.  Obviously $\sqrt{\langle Q\rangle} \rightarrow Z $ for large $Z$.

In Fig. 8 we plot $\sqrt{\langle Q\rangle}$ as a function of $Z$.  
The results are practically identical for $IS$ = (01) and $IS$ = (10).  
Note that $\sqrt{\langle Q\rangle}$ is normalized such as to be identical 
to $Z$ at large $Z$.
One can see that the cluster model gives $\sqrt{\langle Q\rangle}=0$ 
at $Z=0$, consistent with the spherical symmetry of the system, while 
the molecular basis result is $\sqrt{\langle Q\rangle}= 0.573 $ fm at 
$Z=0$, wich suggests that the system acquires a small deformation in 
the molecular basis.  This also means that its r.m.s. radius is larger 
in the molecular basis.

In Figs. 9 and 10 we plot the adiabatic potentials as a function of
$\sqrt{\langle Q\rangle}$ instead of $Z$, for $IS$ = (01) and (10) 
respectively.
As $\sqrt{\langle Q\rangle}\neq 0 $ at any $Z$ in the molecular orbital 
basis, the
corresponding potential is shifted to the right and appears above the
cluster model potential at finite values of $\sqrt{\langle Q\rangle}$ but 
tends
asymptotically to the same value.  The comparison
made in Figs. 9 and 10 is meaningful in the context of a 
Schr\"odinger type equation where the local adiabatic potential
appears in conjunction with an ``effective mass'' depending on 
$\sqrt{\langle Q\rangle}$ also.  However an effective mass can be obtain through RGM 
calculations and our future plan is to perform such calculations.
\section{The middle range attraction}
In principle we expected some attraction at large $Z$ due to the
presence of the Yukawa potential tail in Eq. (23). To see the net
contribution of this part of the quark-quark interaction 
we repeated the calculations in the molecular orbital basis
by completely removing the first term - the Yukawa potential part -
in Eq. (23). The result is shown in Fig. 11. for $SI$= (10).
One can see that beyond $ Z  \approx 1.3 $ fm the contribution
of the Yukawa potential tail is very small, of the 
order of 1-2 MeV.  At small values of $Z$ the Yukawa part of (23)
contributes to increase the adiabatic potential because
it diminishes the attraction in the two body matrix elements.

The missing medium- and long-range attraction can in principle be 
simulated in a simple phenomenological way. For example,
 in Ref. \cite{OK84} this has been achieved at the baryon
level. Here we adopt a more consistent procedure assuming that besides
the pseudoscalar meson exchange interaction of Sec. III there exists
an additional scalar, $\sigma$-meson exchange interaction between quarks.
This is in the spirit of the spontaneous chiral symmetry breaking 
mechanism on which the GBE model is based. The $\sigma$-meson is the 
chiral partner of the pion and it should be considered explicitly.\par

Actually once the one-pion exchange interaction between quarks is admitted,
one can inquire about the role of at least two-pion exchanges.
Recently it was found \cite{RB99} that the two-pion exchange also
plays a significant role in the quark-quark interaction. It enhances the
effect of the isospin dependent spin-spin component of the one-pion
exchange interaction and cancels out its tensor component.
Apart from that it gives rise to a spin independent
central component, which averaged over the isospin wave function of
the nucleon it produces an attractive spin independent interaction.
These findings also support the introduction of a scalar ($\sigma$-meson)
exchange interaction between quarks as an approximate description
of the two-pion exchange loops.\par

For consistency with the parametrization \cite{GL96b} we consider here
a scalar quark-quark interaction of the form
\begin{equation}V_\sigma (r)=
\frac{g_\sigma^2}{4\pi}\frac{1}{12m_i m_j}
\{\theta(r-r'_0)\mu_\sigma^2 \,\frac{e^{-\mu_\sigma r}}{ r}- \frac {4}{\sqrt {\pi}}
\alpha'^3 \exp(-\alpha'^2(r-r'_0)^2)\}.
\end{equation}
where $\mu_\sigma$ = 675 MeV and $r'_0$, $\alpha'$ and the coupling
constant $g^2_\sigma/4\pi$ are arbitrary parameters. In order
to be effective at medium-range separation between nucleons we expect
this interaction to have $r'_0 \neq r_0$ and $\alpha' \neq \alpha$.
Note that the factor $1/m_i m_j$ has only been introduced
for dimensional reasons.\par
We first looked at the baryon spectrum with the same variational 
parameters as before. 
The only 
modification is a shift of the whole spectrum which would correspond 
to taking $V_0 \approx - 60$ MeV in Eq. (21).\par
For the $6q$ system we performed calculations in the molecular basis, 
which is more appropriate than the cluster model basis. We found that
the resulting adiabatic potential is practically insensitive to
changes in $\mu_\sigma$ and $r'_0$ but very sensitive to $\alpha'$.
In Fig. 12 we show results for 
\begin{equation}
r'_0 = 0.86 \, { fm}, ~\alpha'  = 1.47 \, { fm}^{-1},~~ g^2_\sigma/4\pi =
g^2_8/4\pi
\end{equation}
One can see that $V_\sigma$ produces indeed an attractive pocket,
deeper for $SI$ = (10) than for (01), as it should be for the
$NN$ problem. The depth
of the attraction depends essentially on $\alpha'$. The precise values
of the parameters entering Eq. (32) should be determined in 
further RGM calculations.  As mentioned above the 
Born-Oppenheimer potential is in fact the diagonal RGM kernel. 
It is interesting that an attractive pocket is seen in this kernel
when a $\sigma$-meson exchange interaction is combined 
with pseudoscalar meson exchange and OGE interactions (hybrid model),
the whole being fitted to the $NN$ problem \cite{KS99}. 
\section{Summary}
We have calculated the $NN$ potential in the adiabatic approximation 
as a function of $Z$, the separation distance between the centres of
the two $3q$ clusters. 
We used a constituent quark model where quarks interact via pseudoscalar
meson exchange. The orbital part of the six-quark states was constructed either 
from cluster model or molecular orbital single particle states.
The latter are more realistic, having the proper axially and reflectionally
symmetries. Also technically they are more convenient. We explicitly 
showed that they are important at small values of $Z$. In
particular we found that the
$NN$ potential obtained in the molecular orbital basis has a less repulsive
core than the one obtained in the cluster model basis. However 
none of the bases leads to an attractive pocket. We have simulated
this attraction by introducing a $\sigma$-meson exchange interaction
between quarks.\par
To have a better understanding of the two bases we have also calculated 
the quadrupole mement of the $6q$ system as a function of $Z$.  The 
results show that in the molecular orbital basis the system acquires 
some small deformation at $Z=0$.  As a function of the quadrupole 
moment the adiabatic potential looks more repulsive in the molecular 
orbital basis then in the cluster model basis.  In this light one might 
naively expect that the molecular basis will lead to scattering 
phase-shifts having a more repulsive behaviour.\par
The present calculations give us an idea about the size 
and shape of the hard core
produced by the GBE interaction.  Except for small values of $Z$ the 
two bases give rather similar potentials.  Taking $Z$ as a generator 
coordinate the following step is to perform
a dynamical study based on the resonating group method which
will provide phase-shifts to be compared to the experiment. The 
present results constitute an intermediate step towards such a study.\par 
\section{Appendix}
In this appendix we study the behaviour of the confinement potential in the 
molecular orbital basis at large separation distance $Z$ 
between the centres of two $3q$ clusters. 
As an example we consider the state $|42^+[42]_O [33]_{FS}\rangle$.
Through the fractional parentage technique \cite{HA81,ST96} the
six-body matrix elements can be reduced to the calculation
of two-body matrix elements. Using this technique and integrating
in the color space one obtains 
\begin{eqnarray}
&&\langle 42^+[42]_O [33]_{FS} | V_{conf} | 42^+[42]_O [33]_{FS}\rangle =
\frac{1}{40}~ [ 22 \langle \pi\pi |V| \pi\pi\rangle \nonumber \\ 
&&\hspace{3cm}\mbox{}+ 76 \langle \sigma\pi |V| \sigma\pi \rangle 
+ 26 \langle \sigma\pi |V| \pi\sigma\rangle \nonumber \\ 
&&\hspace{3cm}\mbox{}- 58 \langle \pi\pi |V| \sigma\sigma \rangle
+ 22 \langle \sigma\sigma |V| \sigma\sigma\rangle ]
\end{eqnarray}
where the right-hand side contains two-body orbital matrix elements.
According to Eq. (8) for $Z \rightarrow \infty$ one has
\begin{equation}
|\sigma\rangle \rightarrow \frac{1}{\sqrt{2}}~|R~+~L\rangle,\ \ 
|\pi\rangle \rightarrow \frac{1}{\sqrt{2}}~|R~-~L\rangle
\end{equation} 
Replacing these asymptotic forms in the above equation one
obtains matrix elements containing the states $|R\rangle$ and $|L\rangle$.
Most of these martix elements vanish asymptotically. The only
surviving ones are
\begin{equation}
\langle R~R |V| R~R\rangle \rightarrow a~,\,\,
\langle R~L |V| R~L\rangle \rightarrow b~Z
\end{equation}
where $a$ and $b$ are some constants. This brings to the following
asymptotic behaviour of the matrix elements in the right-hand side of (34)
\begin{eqnarray}
\langle \sigma\sigma |V| \sigma\sigma\rangle \rightarrow (a~+~b~Z)/2 \nonumber\\
\langle \pi\pi |V| \pi\pi\rangle \rightarrow (a~+~b~Z)/2  \nonumber\\
\langle \sigma\pi |V| \sigma\pi \rangle \rightarrow (a~+~b~Z)/2 \nonumber\\
\langle \sigma\pi |V| \pi\sigma\rangle \rightarrow (a~-~b~Z)/2 \nonumber\\
\langle \pi\pi |V| \sigma\sigma \rangle \rightarrow (a~-~b~Z)/2
\end{eqnarray}
from which it follows that
\begin{equation}
\langle 42^+[42]_O [33]_{FS} | V_{conf} | 42^+[42]_O [33]_{FS}\rangle 
\rightarrow (11~a~+~19~b~Z)/10
\end{equation}
i.e. this matrix element grows linearly with $Z$ at
large $Z$. 
In a similar manner one can show that the confinement matrix 
element of the state
$\left|{\left.{33{\left[{42}\right]}_{O}{\left[{51}\right]}_{FS}}\right\rangle}
\right.$ the coefficient of the term linear in $Z$ cancels out so that
in this case one obtains a plateau as in Fig. 2.\par 
\vspace{0.8cm}
{\bf{Acknowledgements.}} We are very grateful to L. Wilets, K. Shimizu
and L. Glozman for useful comments.

\begin{table}\label{I}\caption{Parameters of the Hamiltonian (20-25)}
\begin{tabular}{c|c|c|c|c}
$V_0$ (MeV) & $C$ (fm$^{-2})$ & $g_8^2/4 \pi$ &
$g_0^2/4 \pi$ & Reference\\
\hline
0 & 0.474 & 0.67 & 1.206 & \cite{GL96b}\\
\end{tabular}
\end{table}

\begin{table}\label{II}\caption{Variational solution of the Hamiltonian
(20)-(25) for the nucleon mass $m_N$ with $\beta$ as a variational
parameter (see text)}
\begin{tabular}{c|c}
$\beta$ (fm) & $m_N$ (MeV)\\
\hline
0.437 & 969.6\\
\end{tabular}
\end{table}

\newpage
\begin{figure}
\begin{center}
\psfig{figure=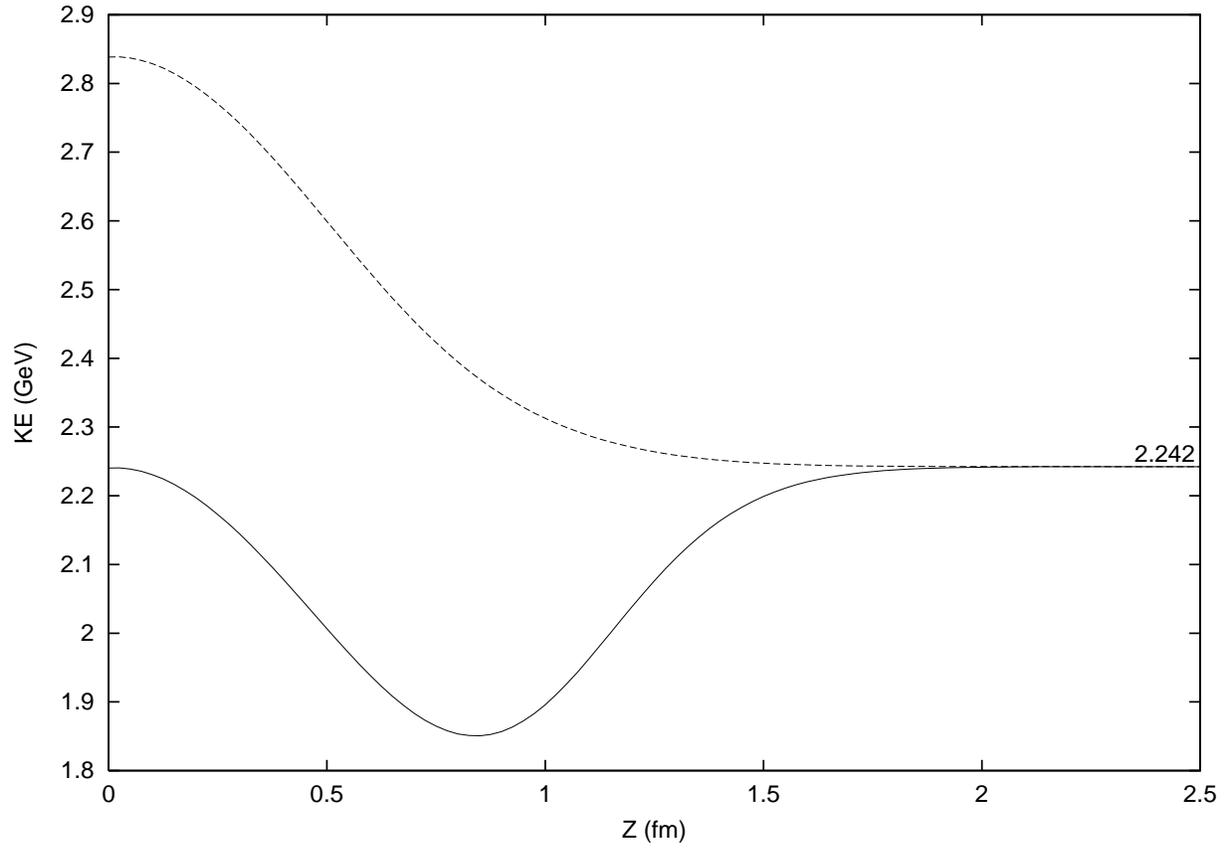,width=16.5cm}
\end{center}
\caption{\label{Fig. 1} The cluster model basis.
The expectation value of the kinetic energy $\langle K.E.\rangle$ 
as a function of the separation distance $Z$ between two $3q$
clusters. The asymptotic value of 2.242 GeV , given by Eq. (28)
is indicated. The full line corresponds to $| [6]_O \rangle$ and the
dashed line to $| [42]_O \rangle $ states.}
\end{figure}

\newpage
\begin{figure}
\begin{center}
\psfig{figure=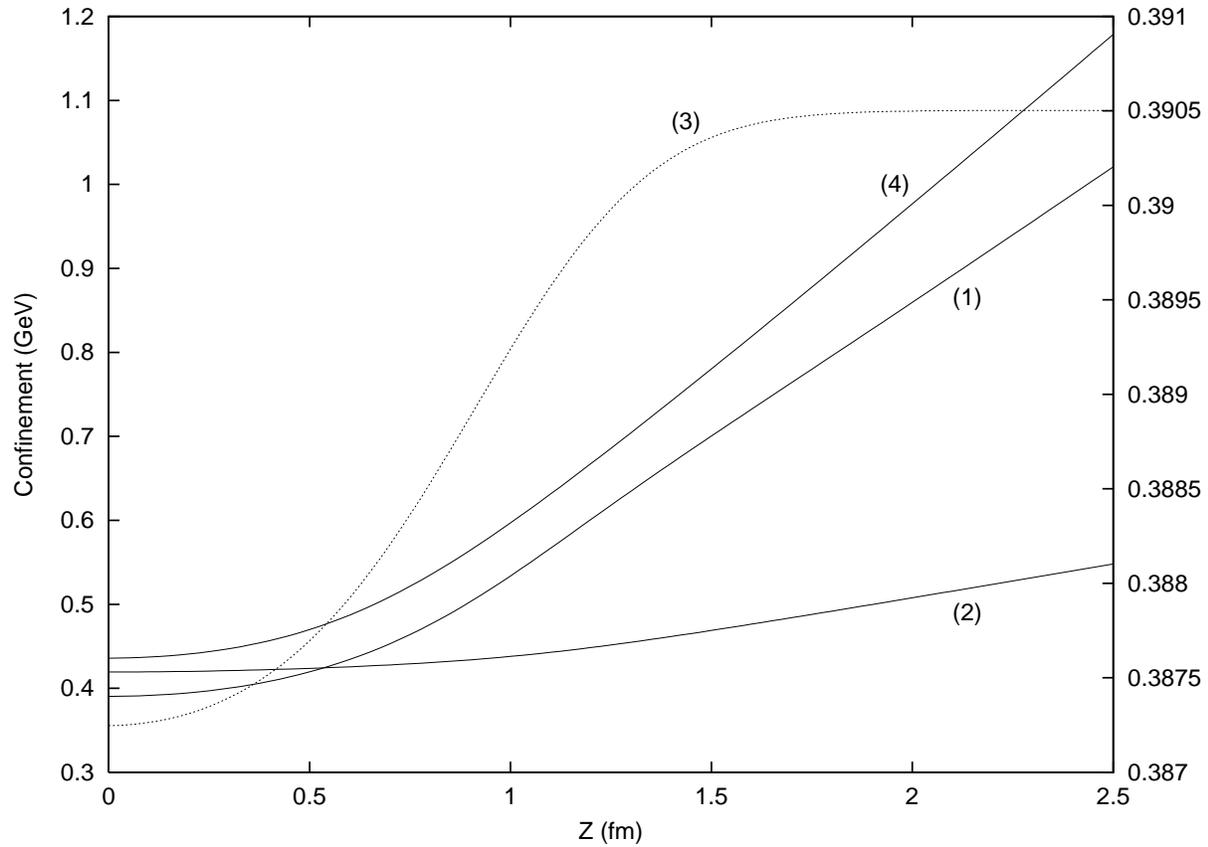,width=16.5cm}
\end{center}
\caption{\label{Fig. 2} The cluster model basis. The expectation value of 
$V_{conf}$ of Eq. (21). The corresponding states are:
(1) - $| [6]_O [33]_{FS} \rangle$,
(2) - $| [42]_O [33]_{FS} \rangle$,
(3) - $| [42]_O [51]_{FS} \rangle$,
(4) - $| [411]_O [51]_{FS} \rangle$.
Note that for curve (3) the scale is on the rhs vertical line.} 
\end{figure}

\newpage
\begin{figure}
\begin{center}
\psfig{figure=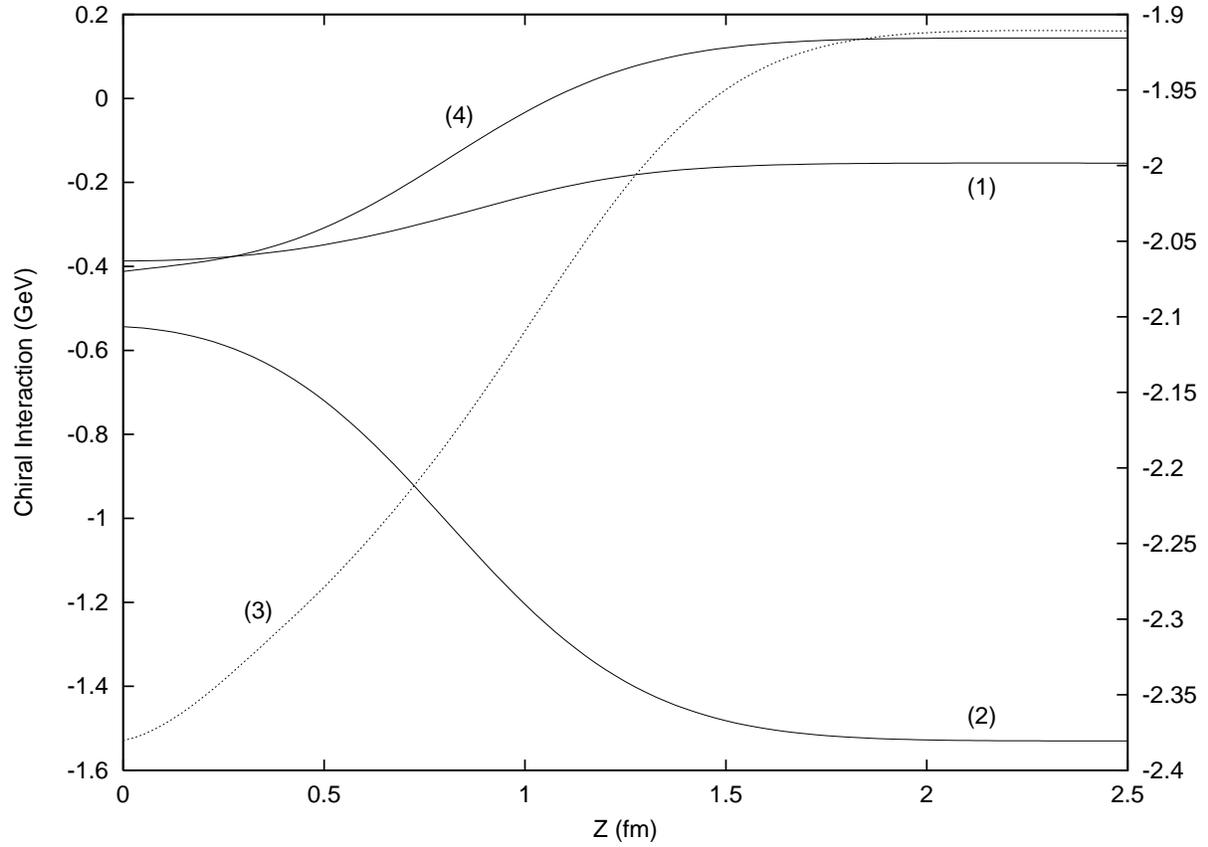,width=16.5cm}
\end{center}
\caption{\label{Fig. 3} The cluster model basis.
The expectation value of the chiral interaction , Eqs. (22)-(25),
for $S$ = 1, $I$ = 0. The curves are numbered as in Fig. 2 and the scale 
for (3) is also on the rhs vertical line.}
\end{figure}

\newpage
\begin{figure}
\begin{center}
\psfig{figure=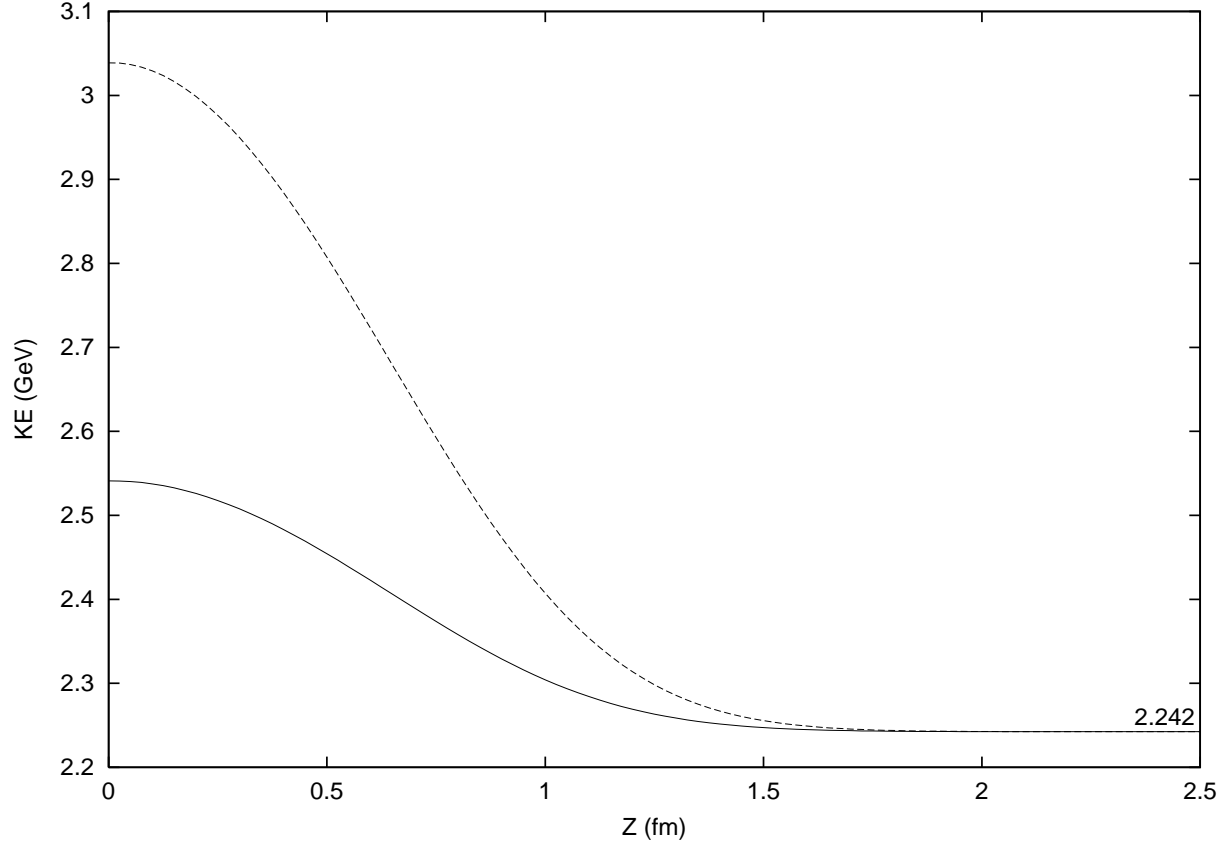,width=16.5cm}
\end{center}
\caption{\label{Fig. 4} The molecular orbital basis.
The expectation value of the kinetic energy $\langle K.E. \rangle$ 
for the $| [6]_O [33]_{FS} \rangle$ (full curve)  and
$| 42^+ [42]_O [51]_{FS} \rangle$ (dashed curve) states (see Eqs. (11) and (17)
respectively). The latter is the most dominant state at $Z$ = 0
(see text).}
\end{figure}

\newpage
\begin{figure}
\begin{center}
\psfig{figure=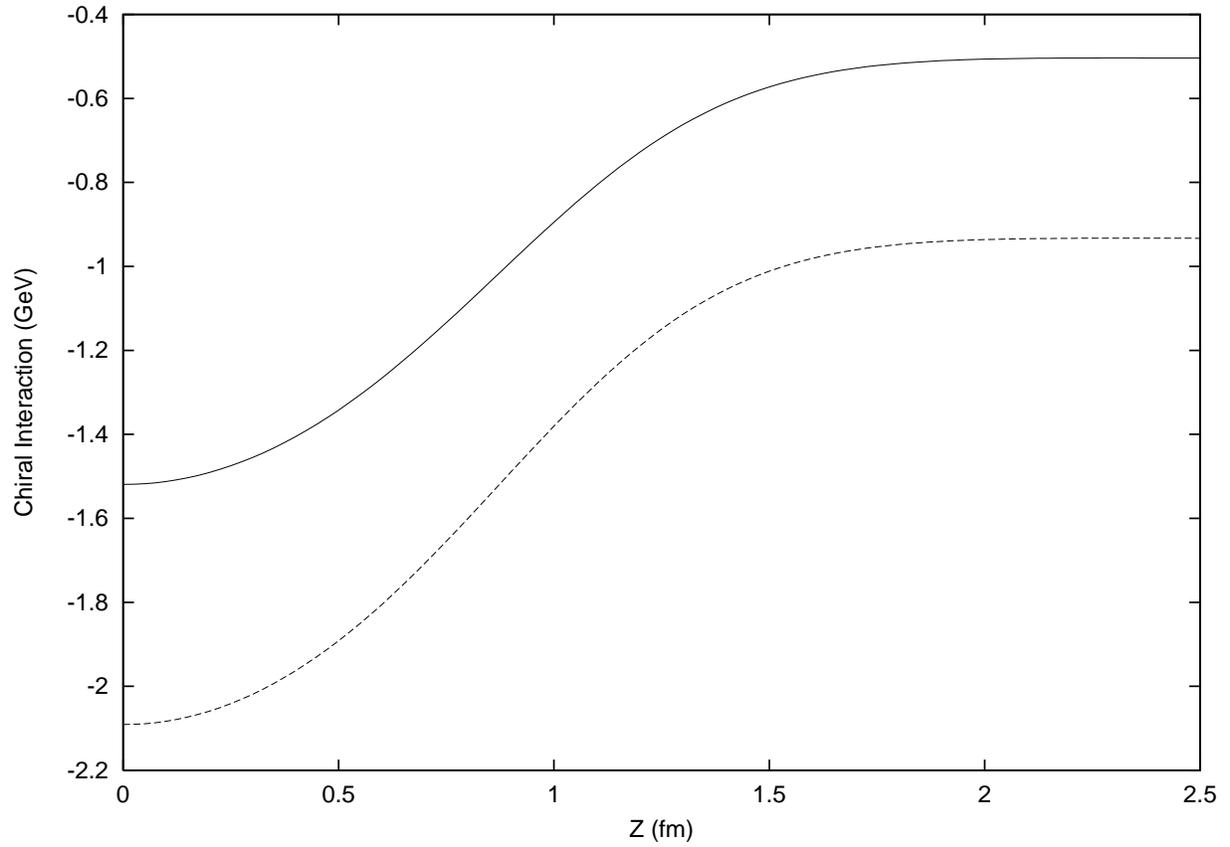,width=16.5cm}
\end{center}
\caption{\label{Fig. 5} The molecular orbital basis.
The expectation value of the chiral interaction, eqs. (22)-(25),
for $| 42^+ [42]_O [51]_{FS} \rangle$ which is
the most dominant state at $Z$ = 0. The dashed curve corresponds to
$S$ = 1, $I$ = 0 and the full curve to $S$ = 0, $I$ = 1.}
\end{figure}

\newpage
\begin{figure}
\begin{center}
\psfig{figure=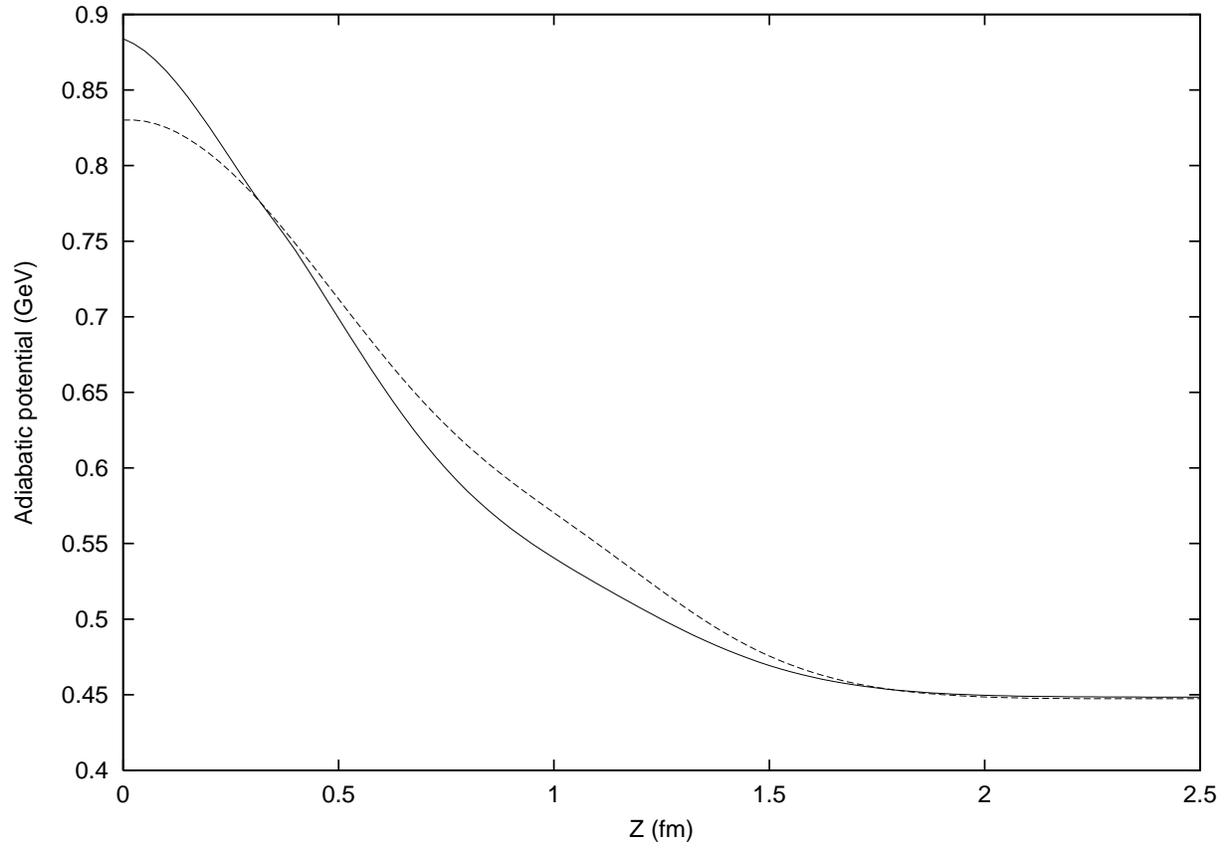,width=16.5cm}
\end{center}
\caption{\label{Fig. 6} Comparison of the adiabatic potential
for $S$ = 1, $I$ = 0, calculated in the cluster model basis (full
curve) and the molecular orbital basis (dashed curve).}
\end{figure}

\newpage
\begin{figure}
\begin{center}
\psfig{figure=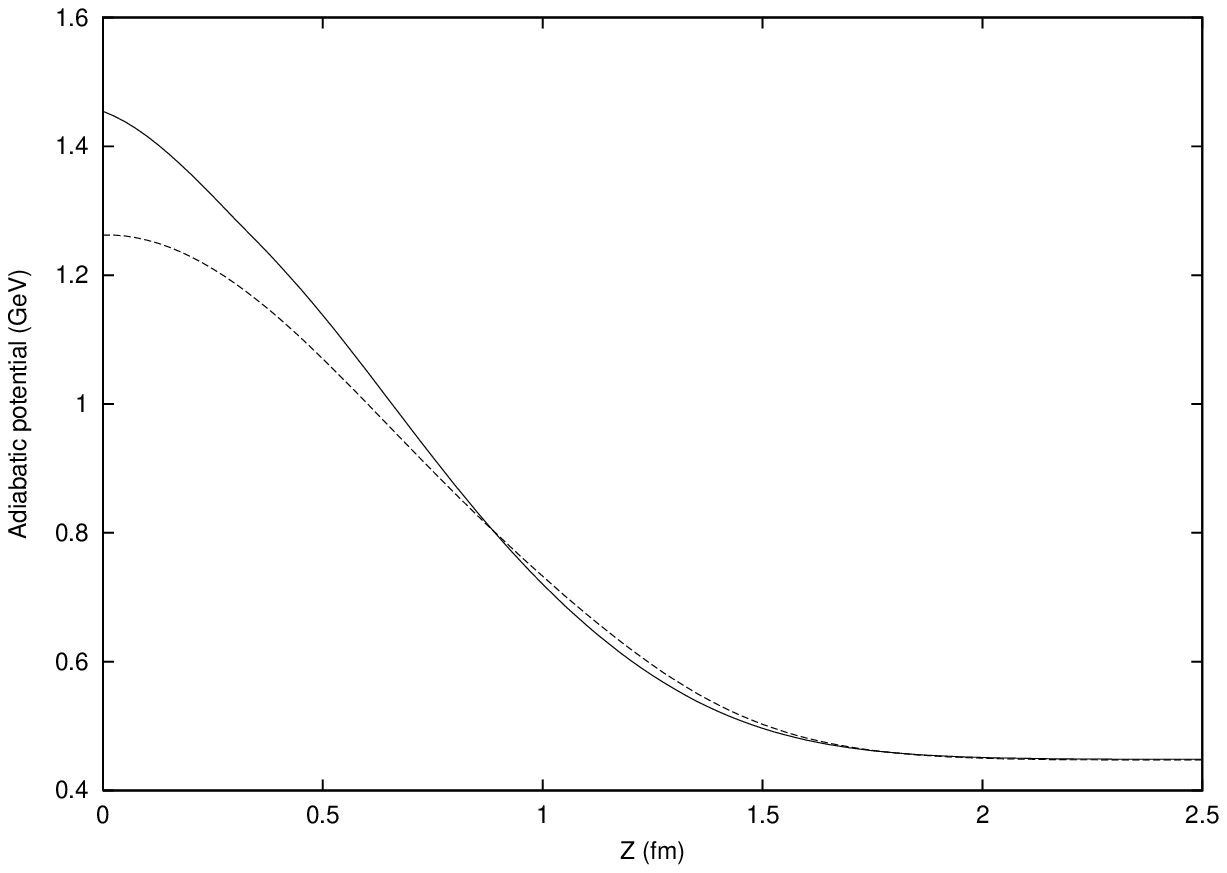,width=16.5cm}
\end{center}
\caption{\label{Fig. 7} Same as Fig. 6 but for $S$ = 0, $I$ = 1.}
\end{figure}

\newpage
\begin{figure}
\begin{center}
\psfig{figure=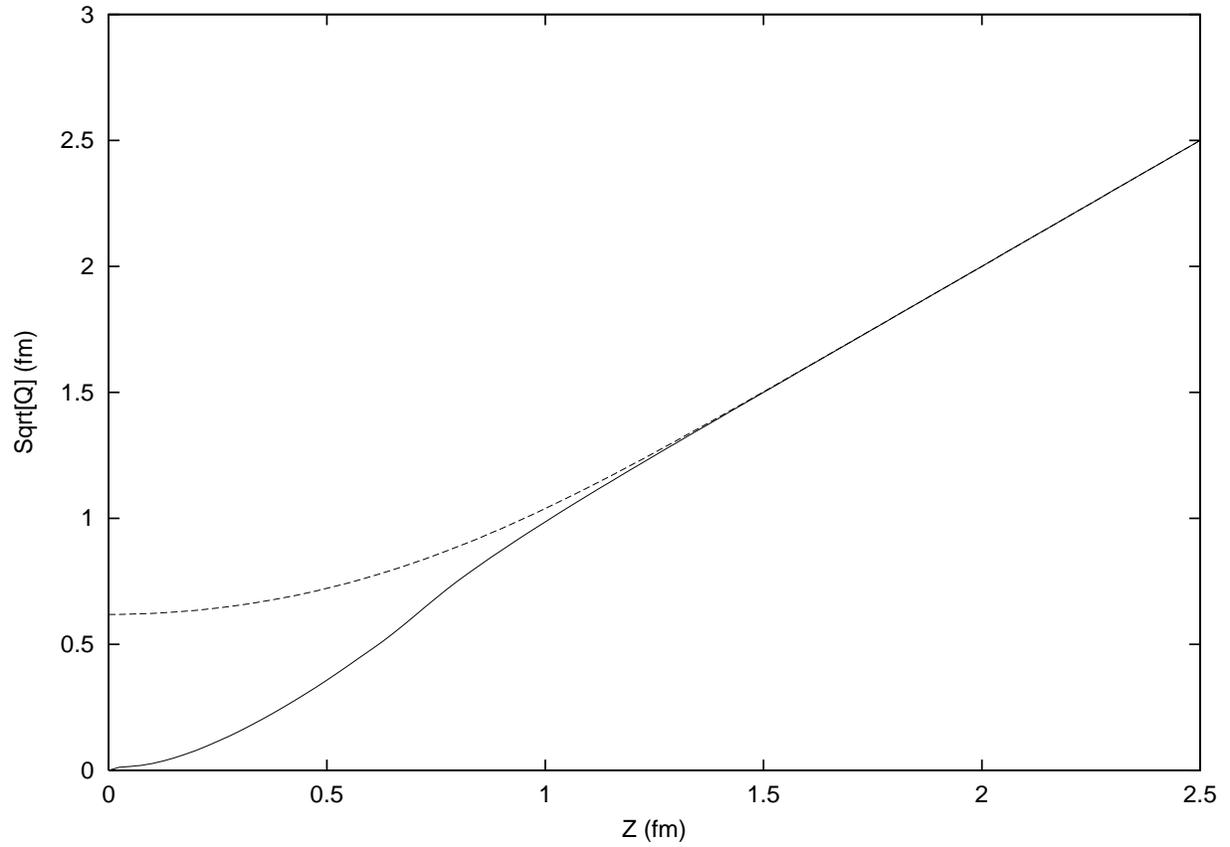,width=16.5cm}
\end{center}
\caption{\label{Fig. 8} $\sqrt{\langle Q\rangle}$ as a function of $Z$ with 
$\langle Q\rangle$ defined by Eq. (31) and normalized as indicated in the 
text.  The full line corresponds to the cluster model basis and the dashed 
line to the molecular orbital basis.}
\end{figure}

\newpage
\begin{figure}
\begin{center}
\psfig{figure=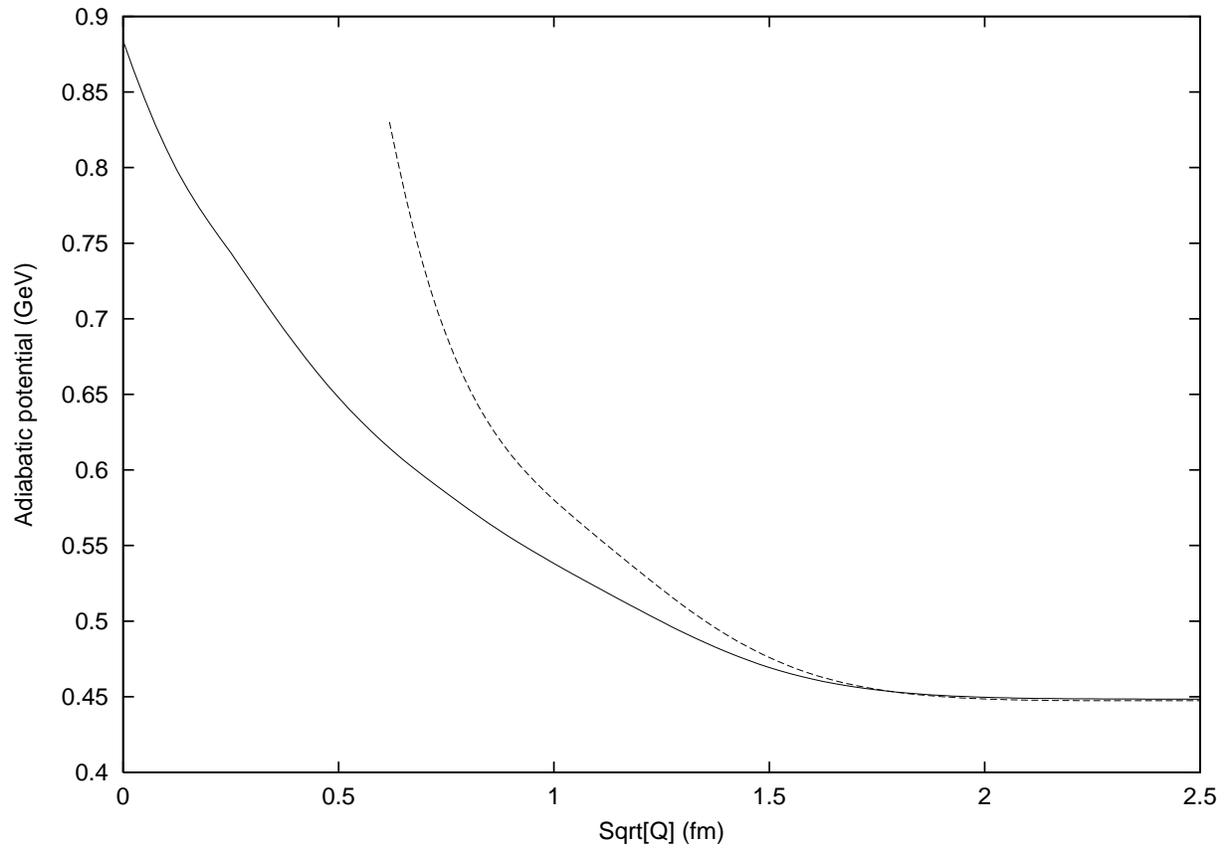,width=16.5cm}
\end{center}
\caption{\label{Fig. 9} Adiabatic potential for $S$ = 1, $I$ = 0 as a function of $\sqrt{\langle Q\rangle}$.  The full line is the cluster model result and the dashed line the molecular basis result.}
\end{figure}

\newpage
\begin{figure}
\begin{center}
\psfig{figure=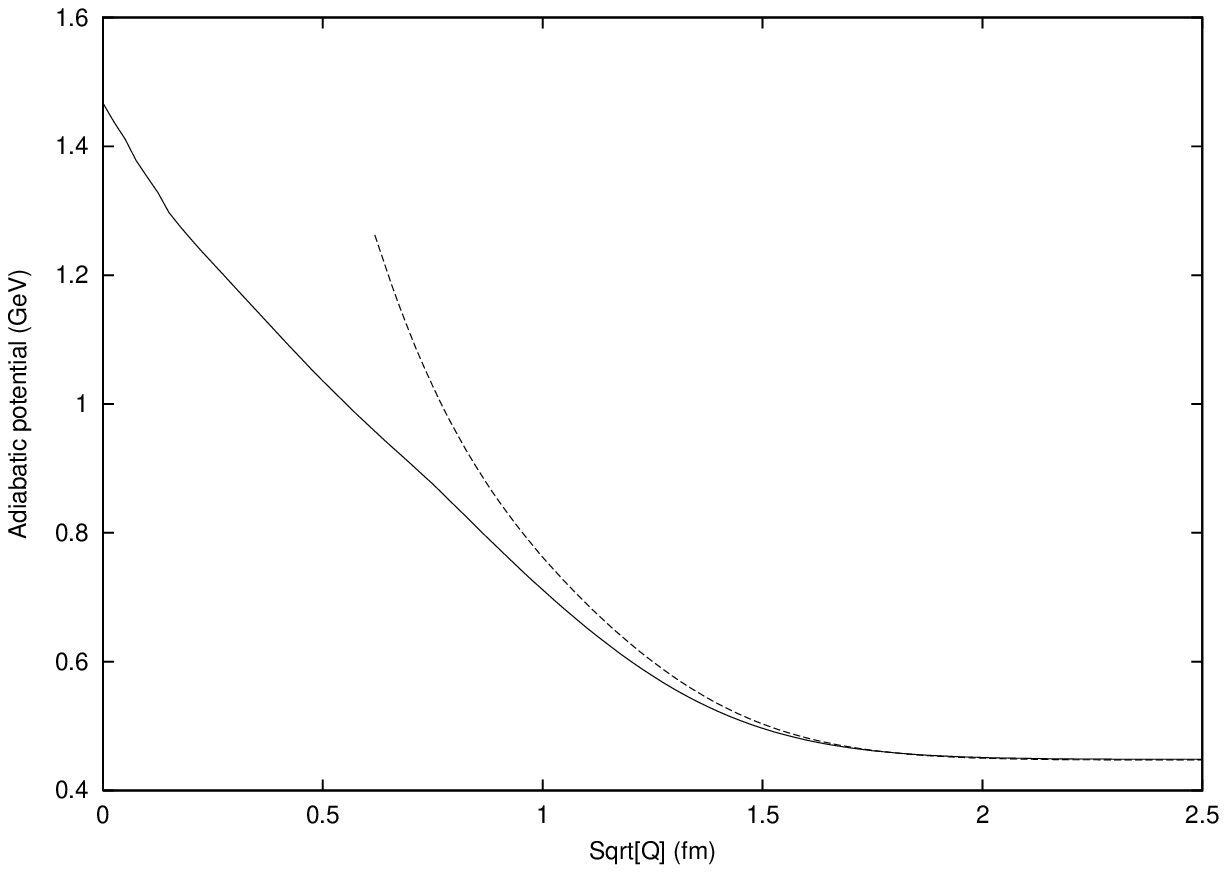,width=16.5cm}
\end{center}
\caption{\label{Fig. 10} Same as Fig. 9 but for $S$ = 0, $I$ = 1.}
\end{figure}

\newpage
\begin{figure}
\begin{center}
\psfig{figure=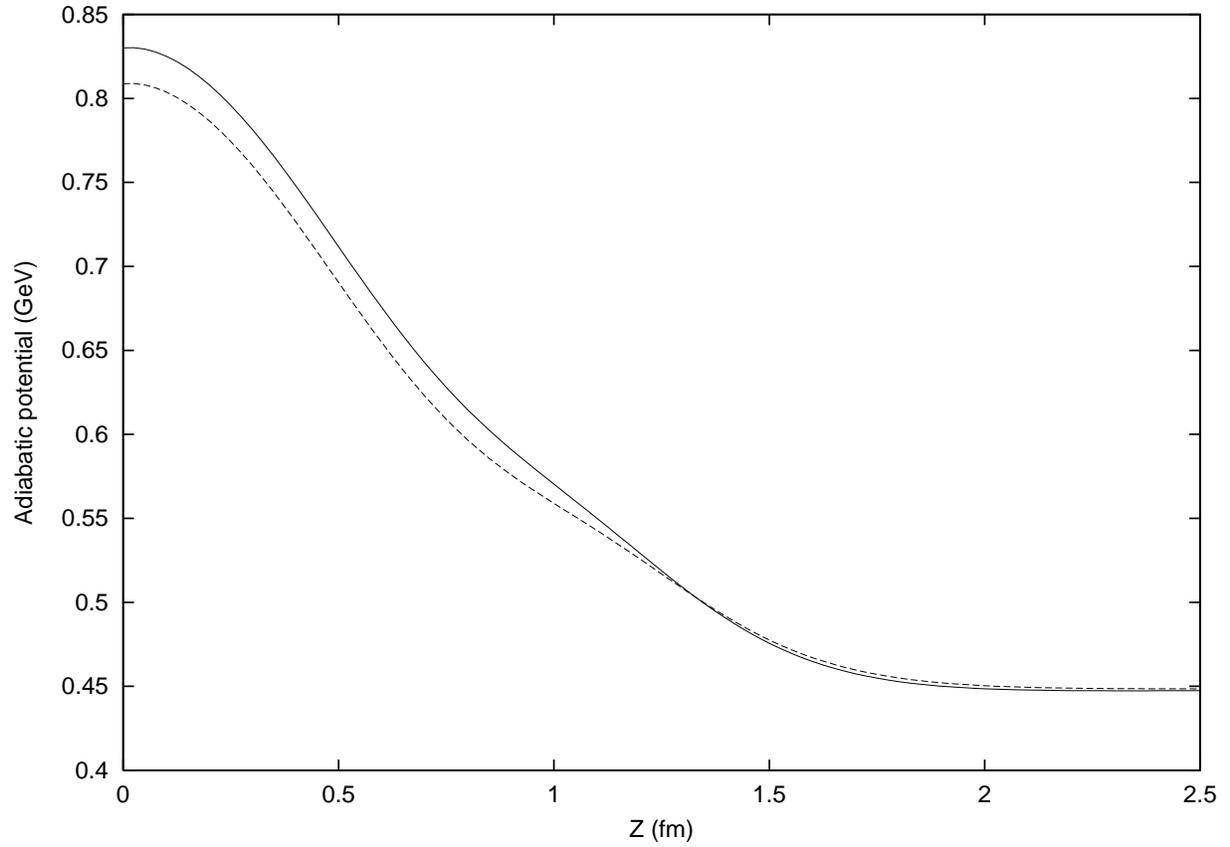,width=16.5cm}
\end{center}
\caption{\label{Fig. 11} The adiabatic potential in the molecular orbital
basis for $SI$ = (10). The solid curve is the same as in Fig. 6.
The dashed curve is the result obtained by removing the 
Yukawa part of the quark-quark interaction (23).}
\end{figure}

\newpage
\begin{figure}
\begin{center}
\psfig{figure=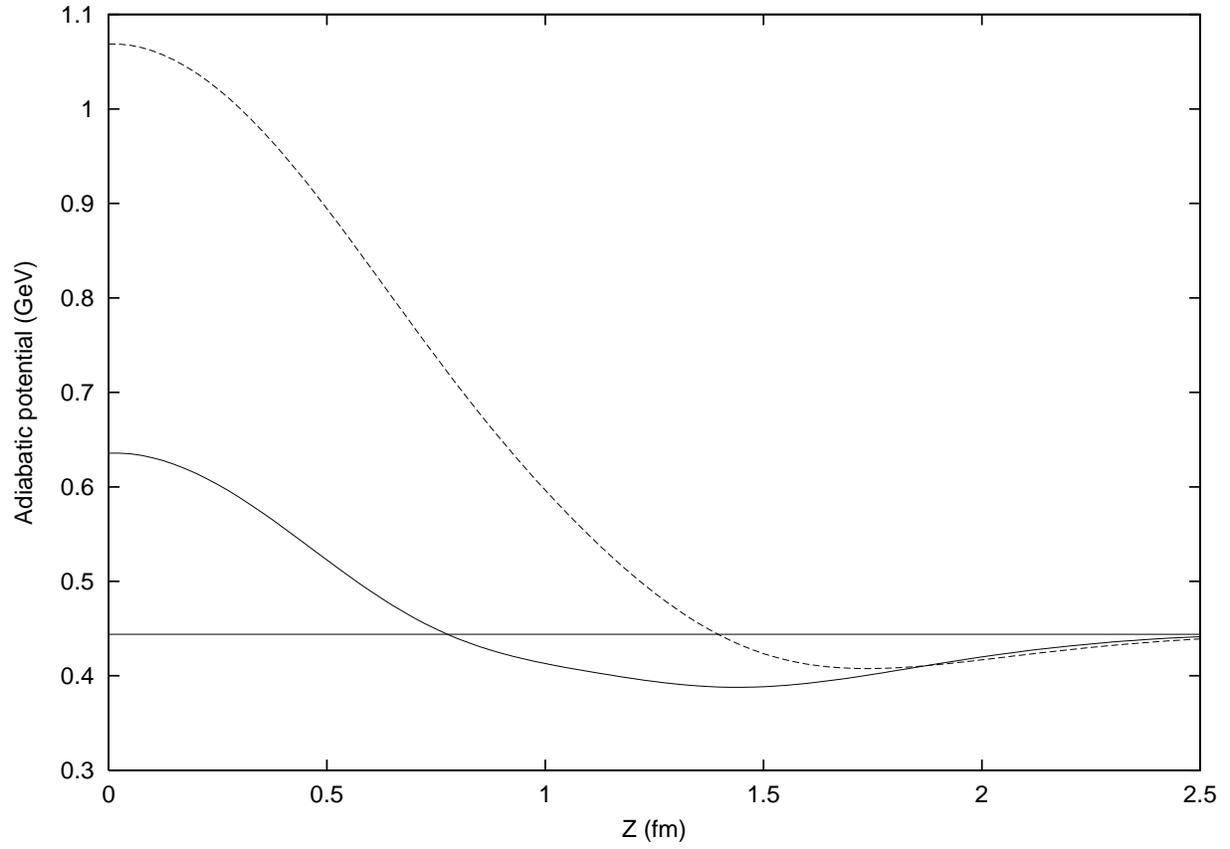,width=16.5cm}
\end{center}
\caption{\label{Fig. 12} The adiabatic potential in the molecular orbital
basis for $SI$ = (10) (full curve) and $SI$ = (01) (dashed curve)
with pseudoscalar + scalar quark-quark interaction.}
\end{figure}

\end{document}